\documentclass[conference,10pt]{IEEEtran}

\usepackage{epsfig}
\usepackage{times}
\usepackage{float}
\usepackage{afterpage}
\usepackage{amsmath}
\usepackage{amstext}
\usepackage{amssymb,bm}
\usepackage{latexsym}
\usepackage{color}
\usepackage{graphicx}
\usepackage{amsmath}
\usepackage{amsthm}
\usepackage{graphicx}
\usepackage[center]{caption}
\usepackage{pstricks}
\usepackage{caption}
\usepackage{subcaption}
\usepackage{booktabs}
\usepackage{multicol}
\usepackage{lipsum}

\allowdisplaybreaks

\def \eu{\mathrm{e}}

\newcommand{\E}{\mathbb{E}}

\newtheorem{thm}{Theorem}
\newtheorem{cor}[thm]{Corollary}
\newtheorem{lem}[thm]{Lemma}

\newtheorem{rem}{Remark}

\title{Entropic CLT for Order Statistics
}%
\author{
\IEEEauthorblockN{Martina Cardone$^*$, Alex Dytso$^{\star}$,  Cynthia Rush$^\dagger$ }
$^*$ University of Minnesota, Minneapolis, MN 55404, USA, Email: mcardone@umn.edu\\
$^{\star}$ New Jersey Institute of Technology, Newark, NJ 07102, USA Email: alex.dytso@njit.edu\\
$^\dagger$ Columbia University, New York, NY 10025, USA, Email: cynthia.rush@columbia.edu
\vspace{-0.9em}
\thanks{Authors are listed in alphabetical order. }
}

\begin{document}
\IEEEoverridecommandlockouts
\maketitle

\begin{abstract}
It is  well known that central order statistics exhibit a central limit behavior and converge to a Gaussian distribution as the sample size grows. This paper strengthens this known result by establishing an entropic version of the CLT that ensures a stronger mode of convergence using the relative entropy. In particular, an order $O(1/\sqrt{n})$ rate of convergence is established under mild conditions on the parent distribution of the sample generating the order statistics. To prove this result, ancillary results on order statistics are derived, which might be of independent interest.
\end{abstract}

\section{Introduction} 
\label{sec:Intro}
Consider a random sample  $X_1, X_2, \ldots, X_n$ drawn independently from a parent distribution having cumulative distribution function (cdf) $F$ and probability density function (pdf) $f$. Let  the random variables $X_{(1)} \le  X_{(2)} \le \ldots \le X_{(n)}$ denote the order statistics of the sample.  With this notation, $X_{(1)}$ corresponds to the minimum value of the sample,  $X_{(n)}$ corresponds to the maximum value of the sample, and  $X_{( \frac{n+1}{2})}$   (provided that $n$ is odd) corresponds to the median of the sample.   Order statistics play an important role in statistical sciences; for example, order statistics are instrumental to constructing a class of robust estimators known as L-estimators.  The interested reader is referred to~\cite{HBS17,David2003Book} for comprehensive summaries on the application and theory of order statistics. 

In this paper, we study the asymptotic behavior of the distribution of the \emph{central} order statistics, that is we are interested in the distribution of  $X_{(pn)}$ for a fixed  $p\in (0,1)$ as $n \to \infty$.
Our main contribution is the proof of
a strong form of the central limit theorem (CLT) showing that the relative entropy between the Gaussian distribution and  the distribution of the order statistics converges to $0$ as $n$ grows.

\subsection{Prior Work}
The study of the asymptotic distribution of order statistics has a long history. For example, in~\cite{laplace1818theorie} Laplace  (already in 1818!) established asymptotic normality of the sample median.  The asymptotic normality of central order statistics was shown by Smirnov in~\cite{smirnov1935uber} and, in~\cite{smirnov1949limit}, Smirnov proved a general convergence theorem establishing that a non-degenerate asymptotic distribution of $X_{(r_n)}$ can be of only  four different types, under the condition that $\sqrt{n} ( \frac{r_n}{n}-p) \to 0 $ for a fixed $p\in (0,1)$.  Further, in~\cite{balkema1978limit, balkema1979limit2}, Balkema and Haan established that the distribution of order statistics is dense in the space of all distributions in the following sense:   there exists a parent cdf such that the limiting cdf of $X_{(r_n)}$  is any desired cdf for an appropriately chosen sequence $r_n$.  A succinct summary of these results can be found in~\cite[p.145]{reiss2012approximate}. 

The asymptotic distribution of the joint order statistics has been studied in a series of papers~\cite{siddiqui1960distribution,weiss1963asymptotic,weiss1979asymptotic,weiss1969asymptotic}.
Berry-Esseen type  results~\cite{feller2008introduction} for the order statistics have been shown by Reiss in~\cite{reiss1974accuracy}. The convergence in the total variational distance has been studied in~\cite{weiss1969asymptotic,ikeda1972uniform,falk1989note} and the law of the iterated logarithm for the order statistics has been shown in~\cite{bahadur1966note}. 
Asymptotic results for random order statistics (i.e., $X_{(\nu)}$ where $\nu$ is an integer valued random variable) have been considered in~\cite{puri2011limit}. 
All of the aforementioned results assume that the parent distribution has an absolutely continuous cdf, while the asymptotic distribution of discrete random variables has recently been considered in~\cite{ma2011asymptotic}.

The entropic CLT for the sample mean has been first studied by Linnik in~\cite{linnik1959information} and later considerably generalized by   Barron in~\cite{barron1986entropy}. There has also been some recent activity around finding the rates of convergence and the interested reader is referred to~\cite{artstein2004rate,johnson2004fisher,madiman2007generalized,bobkov2013rate,bobkov2019renyi} and references therein. 

Information measures on the distribution of order statistics have also received some attention~\cite{baratpour2007some,baratpour2008characterizations,abbasnejad2010renyi, balakrishnan2020cumulative,zheng2009fisher,abbasnejad2010renyi,wong1990entropy,ebrahimi2004information}. For example, in~\cite{dytso2021measuring} we showed that the $f$-divergence between the joint distribution of order
statistics and the product distribution of order statistics does
not depend on the parent distribution.
\subsection{Contributions and Outline}
We establish an entropic version of the CLT for order statistics that  ensures a strong mode of convergence in terms of relative entropy. In particular, in Section~\ref{sec:MainRes} we provide mild conditions to ensure that $D( X_{(np)}  \| G_{n,p}) = O \left( {1}/{\sqrt{n}} \right)$, where $G_{n,p}$ is a Gaussian random variable and $D(\cdot \| \cdot)$ denotes the relative entropy. In order to prove the CLT, in Section~\ref{sec:AncillaryResults} we derive some ancillary results on order statistics, which might be of independent interest. For instance, we show a rather general bound on the moments of the order statistics in terms of properties of the parent distribution. Finally, in Section~\ref{sec:DiscConcl} we conclude the paper with a discussion on the necessity of the derived conditions for convergence. In particular, we show that (although mild) some of the conditions might not be necessary, whereas others (or a variation of them) do appear to be needed to ensure convergence of the relative entropy. Some of the proofs can be found in the appendices.

\subsection{Notation}
Deterministic quantities are denoted by lower case letters and random variables are denoted by upper case letters (e.g., $x,X$). 
The differential entropy of a random variable $V$ having pdf $f_V$ is defined and denoted by
\begin{equation}
h(V) =- \int_{\mathbb{R}} f_V(v) \log f_V(v) \ {\rm d} v.
\label{eq:DiffEntr}
\end{equation} 
The relative entropy between  $W \sim f_W$ and $V \sim f_V$, respectively, is defined and denoted by 
\begin{equation} 
D(W\| V) =\int_{\mathbb{R}} f_W(x) \log \frac{f_W(x)}{f_V(x)} \ {\rm d}x.
\end{equation}   
The inverse cdf, also known as the quantile function, of the cdf $F$ is defined as
\begin{equation}
F^{-1}(p) = \inf \{ x \in \mathbb{R}:  p \le F(x) \}, \, \,\,  p \in (0,1). 
\end{equation} 
The inverse cdf will play an important role in our analysis. In particular, we will often exploit the following well-known fact~\cite{arnold1992first},
\begin{equation}
\label{eq:OrdStatUn}
 X_{(k)} \overset{d}{=} F^{-1}(U_{(k)}), \,\,\, k \in \{1,\ldots, n\},
\end{equation} 
where $ \overset{d}{=}$ denotes equality in distribution and  $U_{(1)}, \ldots, U_{(n)} $ are the order statistics of a sample drawn independently from a  parent distribution that is uniform on $(0,1)$.  We will also frequently use the mean and variance of $U_{(k)}$, which can be characterized by noting that $U_{(k)} \sim \mathsf{Beta}(k, n+1-k)$~\cite{arnold1992first}; hence,  for $k \in \{1,\ldots, n\}$,
\begin{equation}
\E[ U_{(k)} ] = \frac{k}{n+1}, \text{ and }\, {\rm Var}(U_{(k)}) = \frac{k(n+1-k)}{(n+1)(n+2)^2} . \label{eq:Mean_and_variance_Uk}
\end{equation} 
The $L_p$-norm of a function $f$ is denoted as 
\begin{align}
\|f \|_p =\left( \int_{\mathbb{R}} 
|f(x)|^p \ {\rm{d}}x \right)^{ \frac{1}{p}}, \,\,\, p \in [1, \infty],
\end{align}
where $\|f \|_\infty$ is understood as the essential supremum of $f$.  

Throughout the paper, we will use the following notation to describe the limiting behavior of functions as its input tends to infinity. We say that $f(n) = O(g(n))$ for two functions $f$ and $g$ if there exists a $k > 0$ and an $n_0$ such that $f(n) \leq k g(n)$ for all $n > n_0$. Similarly we say that $f(n) = \Theta(g(n))$ if there exists a $k_1, k_2 > 0$ and an $n_0$ such that $ k_1 g(n) \leq f(n) \leq k_2 g(n)$ for all $n > n_0$.

\section{Main Result} 
\label{sec:MainRes}
We begin this section by reviewing the classical CLT for the order statistics due to~\cite{arnold1992first}, and then present our entropic CLT result in Theorem~\ref{thm:EntroVers}. In what follows, we let $G_{n,p} \sim \mathcal{N}( \mu_{p}, V_{n,p} )$ where $( \mu_{p}, V_{n,p} )$ are defined as
\begin{equation}
 \mu_{p}= F^{-1}(p), \qquad \text{ and } \qquad V_{n,p} =  \frac{p (1-p)}{ n  \left( f (F^{-1}(p))  \right)^2}.\label{eq:Var}
\end{equation}
%
%
Consider a random sample  $X_1, X_2, \ldots, X_n$ drawn independently from a parent distribution having cdf $F$ and pdf $f$.
The standard CLT for continuous central order statistics~\cite{arnold1992first} guarantees that for a fixed $p \in (0,1)$,
\begin{equation}
X_{(np)}  \stackrel{d}{\to} G_{n,p},
\end{equation} 
(where $\stackrel{d}{\to}$ indicates convergence in distribution), provided that the following condition holds: $t \mapsto f (F^{-1}(t))$ is continuous at the point $p$ and  $f (F^{-1}(p))>0$. 


 Our main result is that the following theorem, which provides a stronger mode of convergence than the classical CLT for order statistics at the expense of extra sufficient conditions. 
\begin{thm}
\label{thm:EntroVers}
 Let $X_{(1)}, \ldots, X _{(n)}$ be a seqeunce of order statistics generate i.i.d. according to parent cdf $F$ with the pdf $f$, and  let $G_{n,p} \sim \mathcal{N}( \mu_{p}, V_{n,p} )$ where $( \mu_{p}, V_{n,p} )$ are defined in \eqref{eq:Var}. 
Fix some $p \in (0,1)$ and assume that
\begin{enumerate}
\item $\| f\|_m<\infty$ for some $m \in [2,\infty]$;
\item   $f (F^{-1}(p)) \neq 0$,    and    $f'(F^{-1}(t))$ is continuous at $t=p$; and
\item $\E[|X|^r]<\infty$ for some $r>0$. 
\end{enumerate}
Then, 
\[ D( X_{(np)}  \| G_{n,p}) = O \left( {1}/{\sqrt{n}} \right).\]
\end{thm}

\begin{IEEEproof}
We start by recalling the following well known property of the differential entropy (see for example~\cite{rioul2017yet}):  given a differentiable and bijective  function $g$, we have that
\begin{equation}
h\left( g(V) \right) = h(V) +\E \left[  \log | g'(V) | \right].  \label{eq:Mapping_Theorem_for_Entropy} 
\end{equation}
Next,  let $\Phi_{n,p}$ and $\phi_{n,p}$ denote the cdf and the pdf of $G_{n,p}$ (with mean and variance in~\eqref{eq:Var}), and  define a function 
\begin{equation}
g(u) = \Phi_{n,p}( F^{-1} (u)), \,\, \,u\in (0,1). \label{eq:Def_g}
\end{equation} 
The derivative of $g(u)$ is given by 
\begin{equation}
g'(u)\!=\! \frac{\rm d}{{\rm d} u } \Phi_{n,p}( F^{-1} (u)) 
\!=\!    \frac{ \phi_{n,p}( F^{-1} (u)) }{f \left( F^{-1} (u) \right)},   \,u\in (0,1),  \label{eq:gprime}
\end{equation} 
which is strictly positive.  Therefore, $g$ is bijective. 

Let $U$ be the uniform random variable on $(0,1)$. Then,
\begin{equation}
\begin{split}
D( X_{(np)}  \| G_{n,p})& \stackrel{{\rm{(a)}}}{=}   D \left( \Phi_{n,p}(X_{(np)})  \|  \Phi_{n,p} \left( G_{n,p} \right) \right) \\
& \stackrel{{\rm{(b)}}}{=}  D \left( \Phi_{n,p}(X_{(np)})  \|  U  \right)\\
& \stackrel{{\rm{(c)}}}{=}   - h( \Phi_{n,p}(X_{(np)})  )\\
& \stackrel{{\rm{(d)}}}{=} - h( \Phi_{n,p}( F^{-1} (U_{(np)}))\\
& \stackrel{{\rm{(e)}}}{=}    - h( U_{(np)}) -\E[ \log |g'(U_{(np)})| ], \label{eq:thm_step3}
\end{split}
\end{equation}
where the labeled equalities follow because: ${\rm{(a)}}$ $D(W\| U) = D(f(W) \| f(U))$ for any invertible function $f$;  
${\rm{(b)}}$ $F_X(X) \stackrel{d}{=} U$ for any random variable $X$ with cdf $F_X$; 
${\rm{(c)}}$ $D(X\|U)= - h(X)$ for any random variable $X \in (0,1)$; 
$\rm{(d)}$~\eqref{eq:OrdStatUn};
and ${\rm{(e)}}$~\eqref{eq:Mapping_Theorem_for_Entropy} and the definition of $g$ in~\eqref{eq:Def_g}, which is bijective. 

Now we study the terms on the right-hand side of~\eqref{eq:thm_step3}. 
In particular, we start focusing on the term $\E[ \log |g'(U_{(np)})| ]$. 
First, notice that since $\phi_{n,p}(\cdot)$ is the Gaussian pdf having mean and variance given in~\eqref{eq:Var}, we have
\[ \phi_{n,p}( F^{-1} (u)) = \frac{\exp\left(-\frac{1}{2 V_{n,p}}\left(F^{-1} (u) - F^{-1} (p)\right)^2\right)}{\sqrt{2 \pi V_{n,p}}} .\]
Therefore, using~\eqref{eq:gprime} and the above, we obtain
\begin{align}
&\E[ \log |g'(U_{(np)})| ]  \notag\\
& =\E \left[ \log \left( \phi_{n,p}( F^{-1} (U_{(np)}))   \frac{1}{f ( F^{-1} (U_{(np)}) )}   \right)  \right] \nonumber \\
& = -\frac{1}{2} \log \left( 2 \pi V_{n,p} \right)    -  \frac{1}{2 V_{n,p}}  \E \left[ \left( F^{-1} (U_{(np)}) -  F^{-1} (p)   \right)^2\right] \notag\\
& \qquad   - \E \left[ \log \left( f( F^{-1} (U_{(np)}))\right)  \right] \nonumber \\
& = -\frac{1}{2}\log \left( \frac{2 \pi \eu p  (1-p)}{n}  \right)  + \E \left[ \log \left( \frac{f ( F^{-1} (p) )}{f( F^{-1} (U_{(np)}) )}\right)  \right] \notag\\
&\qquad +\frac{1}{2}  -  \frac{ 1}{2 V_{n,p}} \E \left[ \left( F^{-1} (U_{(np)}) -  F^{-1} (p)   \right)^2 \right],
\label{eq:thm_step4}
\end{align}
where the last equality follows by using the expression of $V_{n,p}$ in~\eqref{eq:Var}.
Now, combining~\eqref{eq:thm_step3} and~\eqref{eq:thm_step4}, we find that
\begin{subequations}
\label{eq:IntStep}
\begin{equation}
 D( X_{(np)}  \| G_{n,p}) = K_1+ K_2+ K_3, 
\end{equation}
where
\begin{align} 
K_1 &= \frac{1}{2}\log \left( \frac{2 \pi \eu p  (1-p)}{n}  \right)  -h( U_{(np)}), \label{eq:K1}\\
K_2&=  \frac{1}{2 V_{n,p}} \E \left[ \left( F^{-1} (U_{(np)}) -  F^{-1} (p)   \right)^2 \right] -\frac{1}{2}, \label{eq:K2} \\
K_3&=     \E \left[ \log \left( \frac{f( F^{-1} (U_{(np)}) )}{f ( F^{-1} (p) )}\right)  \right] . \label{eq:K3}
\end{align}
\end{subequations}
It is worth noting that $K_1$ is independent of the parent distribution, whereas the terms $K_2$ and $K_3$ both depend on the parent distribution.
In Section~\ref{sec:AncillaryResults}, we will present ancillary results that show the following facts:
(i) $K_1=O ({1}/{n})$ by Lemma~\ref{lem:uniform_entropy};
(ii) $K_2=O ({1}/{\sqrt{n}})$ by Corollary~\ref{cor:T2asymptotic} (this requires Assumptions \textbf{(2)} and \textbf{(3)});
and (iii) $K_3 = O( {1}/{\sqrt{n}} )$ by Lemma~\ref{lem:Log_Term_bound}  (this requires Assumptions \textbf{(1)} and \textbf{(2)}).
Combing these facts together with~\eqref{eq:IntStep} and the non-negativity property of the relative entropy, we have that 
\begin{equation*}
D( X_{(np)}  \| G_{n,p}) = O ( {1}/{\sqrt{n}} ).
\end{equation*}
This concludes the proof of Theorem~\ref{thm:EntroVers}.

%
%
%
%
%
%
%
\end{IEEEproof}

We conclude this section by highlighting that a variety of different distributions (e.g., uniform, Gaussian, exponential) satisfy the conditions of Theorem~\ref{thm:EntroVers}.  Moreover, it is also interesting to note that the Cauchy distribution satisfies the conditions despite the fact that in this case $\E[|X|]=\infty$; hence, the CLT for the \emph{sample mean} does not hold. This can be seen since the Cauchy pdf clearly satisfies the first two conditions and $\E[|X|^r]<\infty$ for $ 0<r<1$.

\section{Ancillary Results} 
\label{sec:AncillaryResults}
In this section we present auxiliary results needed for the proof of Theorem~\ref{thm:EntroVers}, some of which may be of interest on their own.

\subsection{Entropy of Uniform Order Statistics}
\label{sec:Entropy_unif} 
We provide the exact expression for the differential entropy of $U_{(k)}$ for $k \in \{1,\ldots, n\}$, and an asymptotic expression for the entropy of $U_{(pn)}$ for $p \in (0,1)$ as $n \to \infty$. 
The proof of Lemma~\ref{lem:uniform_entropy} can be found in Appendix~\ref{sec:Limit1}.
\begin{lem}
\label{lem:uniform_entropy}
For any $k \in  \{1, 2, \ldots, n\}$,
\begin{equation}
\label{eq:entrUk}
h(U_{(k)})=   T_{k-1} +T_{n-k}-T_n  -H_n,
\end{equation} 
where for $r \in \mathbb{N}$, with $\mathbb{N}$ denoting the natural numbers,
\begin{align}
H_r &=\sum_{k=1}^r \frac{1}{k}, \qquad \text{ and } \qquad T_r = \log(r!) - r H_r. 
\label{eq:harmonic}
\end{align}
Moreover,   if $k=pn$ for $p \in (0,1)$, then 
\begin{align}
& h(U_{(np)})- \frac{1}{2} \log \left(2\pi \eu \frac{p (1-p)}{n}\right)  \notag\\
 &\hspace{1.5cm}  =   \frac{\frac{1}{ p}+\frac{1}{ 1-p}-4}{6n}+\frac{1}{12n^2}
+O \left( \frac{1}{n^3} \right).\label{eq:approximation_of_h(u)}
\end{align} 
\end{lem} 

It is interesting to note that in Lemma~\ref{lem:uniform_entropy}, the rate of convergence is $O(1/n^2)$ instead of $O(1/n)$ for $p=1/2$.

\subsection{Bound on the Estimation Error of the $p$-th Quantile} 

In practice, one might desire to estimate the $p$-th quantile $F^{-1}(p)$ of an unknown cdf $F$.
The order statistic $X_{(np)}$ based on an i.i.d. sample from the parent cdf $F$ is a natural estimate for  $F^{-1}(p)$. 
It is well known that this estimator is consistent as $n \to\infty$~\cite{arnold1992first}. The next result (see Appendix~\ref{sec:Limit2} for the proof) provides an upper bound on the mean squared error of estimating $F^{-1}(p)$ with $X_{(np)}$. 
\begin{lem}
\label{lem:term_T2}
Fix some $p \in (0,1)$, and assume a pdf $f$ such that $f (F^{-1}(p)) \neq 0$,    and    $f'(F^{-1}(t))$ is continuous at $t=p$. Then, for any  $\epsilon\in (\frac{p}{n+1}, p)$, we have that 
\begin{subequations}
\begin{align}
&\E \left [ \left( F^{-1} (U_{(np)}) -  F^{-1} (p)   \right)^2 \right]   \notag\\ 
&  \hspace{.2cm}  \leq   4       \sqrt{ \E \left[ (X_{(np)})^4 \right]   }    \eu^{  - (n+2)  \left(\epsilon -\frac{p}{n+1} \right)^2 } + \frac{  {\rm Var}(U_{(np)})}{ \left(  f (F^{-1}(p))  \right)^{2} }\notag\\
&  \hspace{.2cm}  \quad + C_{p,\epsilon}  \, O \left( {1}/{n^\frac{3}{2}} \right) , \label{lem:termT2_1}
  \end{align}
  where 
  \begin{equation}
   C_{p,\epsilon}
   =  \max_{ t\in [ p-\epsilon, \, p+\epsilon]}   \left|   \frac{  f'(F^{-1}(t))  }{  ( f (F^{-1}(t))  )^3 } \right| .
  \end{equation} 
  \end{subequations}
\end{lem}

\begin{rem} 
\label{rem:Uni}
Note that if the parent distribution is the uniform on $(0,1)$, then  $F^{-1}(x) = x$ for $x \in (0,1)$ and 
\begin{align*}
 \E  [ ( F^{-1} (U_{(np)}) -  F^{-1} (p)   )^2 ] & = \E \left [ \left (U_{(np)} - p \right )^2\right ] \\
& =  {\rm Var}(U_{(np)})+  \left(\E[ U_{(np)}] -p   \right)^2
\\& = {\rm Var}(U_{(np)})+\frac{p^2}{ (n+1)^2},
\end{align*}
where we have used that $\E[ U_{(np)}]=\frac{np}{n+1}$ (see~\eqref{eq:Mean_and_variance_Uk}) so that $\left (\E[ U_{(np)}] - p   \right )^2  =  (\frac{p}{n+1})^2$.
In other words, there exists a distribution for which the bound in Lemma~\ref{lem:term_T2} is asymptotically tight as $n \to \infty$.

\end{rem} 

\subsection{New Bound on the Moments of Order Statistics} 
In order to control the error in Lemma~\ref{lem:term_T2}  we need to control the fourth moment of the order statistics. Ideally, we would like to control this in terms of the properties of the parent distribution (e.g., moments).  The next result, the proof of which is in Appendix~\ref{app:BOund_On_Moments}, establishes a rather general bound on the moments of the order statistics in terms of properties of the parent distribution. 
\begin{lem} \label{lem:BOund_On_Moments} Let $X_1, X_2, \ldots, X_n$ be an i.i.d.\ random sample. For any $q,r>0$ and $k \in \{1, 2, \ldots, n\}$, we have that
\begin{equation*}
 \E [|X_{(k)}|^q]  \le  C_{n,k,q,r}     \left( \E [|X|^r] \right)^{\frac{q}{r}},
 \end{equation*} 
 where
 \begin{align*}
  &C_{n,k,q,r} \notag\\  
&=   \left \{  \hspace{-0.15cm} \begin{array}{cc}
 \frac{ \Gamma(n+1) \Gamma\left(k-\frac{q}{r} \right) \Gamma\left(n-k-\frac{q}{r}+1 \right)}{\Gamma \left(n-\frac{2q}{r}+1 \right) \Gamma(k)\Gamma(n-k+1)} &    \text{if } k> \frac{q}{r}, \,\, n-k >\frac{q}{r}-1,\\
 \infty & \text{ otherwise}.
 \end{array}  \right.  
 \end{align*} 
 In addition,
 \begin{equation}
 \lim_{n \to \infty}     C_{n,pn,q,r} =    \left( p (1-p) \right)^{  \frac{q}{r} }.  \label{eq:limit_of_C}
 \end{equation} 
\end{lem}

Notice that Lemma~\ref{lem:BOund_On_Moments} shows that all of the moments of the order statistics exist, provided that a single moment of the parent distribution, for any order $r>0$, exists. 

We now conclude this subsection with the following corollary, which is an immediate consequence of Lemma~\ref{lem:term_T2} and Lemma~\ref{lem:BOund_On_Moments} and is used in the proof of Theorem~\ref{thm:EntroVers} to show that $K_2 = O(1/\sqrt{n})$.
\begin{cor}\label{cor:T2asymptotic} Fix some $p \in (0,1)$. Suppose that  $f (F^{-1}(p)) \neq 0$,        $f'(F^{-1}(t))$ is continuous at $t=p$, and  $\E[|X|^r]<\infty$ for some $r>0$. 
   Then,  for $V_{n,p}$ defined in \eqref{eq:Var},
\begin{equation*}
   \frac{1}{2 V_{n,p}}\E \left[ \left( F^{-1} (U_{(np)}) -  F^{-1} (p)   \right)^2 \right]  -\frac{1}{2}   = O \left( \frac{1}{\sqrt{n}} \right). 
   \end{equation*}
\end{cor}

\begin{IEEEproof}
We have that
\begin{align*}
& \frac{1}{2 V_{n,p}} \E \left[ \left( F^{-1} (U_{(np)}) -  F^{-1} (p)   \right)^2 \right]  - \frac{1}{2} \notag \\
 & \stackrel{{\rm{(a)}}}{\le}   \frac{  {\rm Var}(U_{(np)})}{  2 V_{n,p} \left(  f (F^{-1}(p))  \right)^{2} }-\frac{1}{2}+O \left( \frac{1}{\sqrt{n}} \right) \notag \\
  & \stackrel{{\rm{(b)}}}{=}    \frac{n^2(n+1-np)}{ 2(1-p)   (n+1)(n+2)^2} -\frac{1}{2} +O \left( \frac{1}{\sqrt{n}} \right) \notag \\
  & \le     \frac{n+1-np}{ 2(1-p)   n} -\frac{1}{2} +O \left( \frac{1}{\sqrt{n}} \right) \notag \\
   & =     \frac{1}{2 (1-p)   n}  +O \left( \frac{1}{\sqrt{n}} \right),
   \end{align*}
   where ${\rm{(a)}}$ follows by using Lemma~\ref{lem:term_T2} and Lemma~\ref{lem:BOund_On_Moments} since
   \[  \frac{2  \sqrt{ C_{n,np,4,r}  } }{ V_{n,p}}       \left( \E [|X|^r] \right)^{\frac{2}{r}}     \eu^{  - (n+2)  \left(\epsilon -\frac{p}{n+1} \right)^2 } 
 +  \frac{C_{p,\epsilon} }{2 V_{n,p}} \, O \left( \frac{1}{n^\frac{3}{2}} \right)\] 
 is $O( {1}/{\sqrt{n}})$  and ${\rm{(b)}}$ follows from~\eqref{eq:Mean_and_variance_Uk} and~\eqref{eq:Var}.
\end{IEEEproof} 

\subsection{Bound on $K_3$ in~\eqref{eq:IntStep}}
We now conclude this section with the following lemma, the proof of which is given in Appendix~\ref{app:lem:Log_Term_bound}. In particular, Lemma~\ref{lem:Log_Term_bound} demonstrates that $K_3 = O ({1}/{\sqrt{n}})$ as long as $\| f\|_m$ has a finite norm for any $m\geq2$ including $m=\infty$. 
\begin{lem}\label{lem:Log_Term_bound} Fix some $p\in (0,1)$.  Choose some  $q ,r \in [1,\infty]$ such that $\frac{1}{q}+\frac{1}{r}=1$ and $\epsilon>0$ such that 
\begin{equation*}
p>\epsilon> \max  \left\{  \frac{p}{n+1},  \frac{|(q-2) p-q +1|}{q(n-1)+2}  \right \} .
\end{equation*} 
Then,
\begin{align*}
& \E \left[ \log \left( \frac{f ( F^{-1} (U_{(np)}) )}{f ( F^{-1} (p) )}\right)  \right]  \notag\\
& \le 2  \lvert \log ( f ( F^{-1} (p )))  \lvert    \eu^{-2 (n+2)  \left(  \epsilon-\frac{p}{n+1} \right)^2 }+C_\epsilon^{(2)} O \left( \frac{1}{\sqrt{n}} \right) \notag\\
&\quad + 2 C_q \left( \| f \|_{r+1}\right )^{\frac{r+1}{r}} n^{\frac{1}{2}\left(1 - \frac{1}{q}\right)} \eu^{- 2(n+2) \left( \epsilon- \frac{|(q-2) p-q +1|}{q(n-1)+2}  \right)^2   },
\end{align*}
where   the universal constant $C_q := {\rm{e}}^{1+\frac{2}{q}} \left(\sqrt{2 \pi }\right)^{\frac{1}{q}-1}  q^{- \frac{1}{2q}}$ and 
\begin{align*}
C^{(2)}_\epsilon= \max_{ p-\epsilon \le u \le p+\epsilon}  \left| \frac{f' \left( F^{-1} \left(u \right) \right) }{ \left( f \left( F^{-1} \left( u\right)  \right) \right)^2 }  \right|;
\end{align*} 
\end{lem}

\section{Discussion and Conclusion}
\label{sec:DiscConcl}
In this paper, we have derived an entropic version of the CLT for the order statistics that ensures a stronger mode of convergence (in terms of relative entropy) than the convergence in distribution provided by the classical CLT for continuous central order statistics~\cite{arnold1992first}.

The three sufficient conditions in Theorem~\ref{thm:EntroVers} are fairly mild and we discuss them now in more detail.
We suspect that condition 3) (or a variation of it) in Theorem~\ref{thm:EntroVers} might be needed for convergence. To support this claim, consider the following density
\begin{equation}
\label{eq:Den2}
f_1(x)= \frac{2}{x \log^3(x)},  \quad x\in ( \eu, \infty),
\end{equation}
which is a canonical example of a pdf such that $\E[|X|^r] =\infty$ for all $r>0$ (i.e., $f_1$ does not satisfy the third condition in Theorem~\ref{thm:EntroVers}). 
In Appendix~\ref{app:examples}, we indeed show that for this density $K_2 = \infty$; hence, $D( X_{(np)}  \| G_{n,p}) = \infty$.

Unlike condition 3), condition 1) in Theorem~\ref{thm:EntroVers} might not be needed, as we argue next. Consider the following density,
\begin{equation}
f_2(x)= \frac{1}{x \log^2(x)}, \quad x\in (0, \eu^{-1}), \label{eq:Den1}
\end{equation} 
which is a canonical example of pdf such that $\|  f_2\|_m=\infty$ for all $m>1$ (i.e., $f_2$ does not satisfy the first condition in Theorem~\ref{thm:EntroVers}).
Although this density does not satisfy the first condition in Theorem~\ref{thm:EntroVers}, in Appendix~\ref{app:examples} we show that 
\begin{equation*}
D( X_{(np)}  \| G_{n,p}) =\Theta \left( \frac{1}{n} \right),
\end{equation*}
that is, the claim of Theorem~\ref{thm:EntroVers} still holds.
This example shows that, even if pretty mild, condition 1) in Theorem~\ref{thm:EntroVers} might not be needed for convergence. Therefore, an interesting research direction would consist of further relaxing this condition.

\begin{appendices}

\section{Proof of Lemma~\ref{lem:uniform_entropy}}
\label{sec:Limit1}
We first recall that by \cite{arnold1992first}, the random variable $U_{(k)}$ is distributed as $\mathsf{Beta}(k, n+1-k)$; hence, it has pdf
\begin{align}
\label{eq:U_pdf}
f_{U_{(k)}}(x)= c_k x^{k-1}  (1-x)^{n-k}, \qquad \text{ for } x \in [0,1], 
\end{align}
where $c_k=\frac{n!}{ (k-1)!  (n-k)!}$.
Then, using the definition of differential entropy in~\eqref{eq:DiffEntr}, we obtain
\begin{equation}
\begin{split}
-h(U_{(k)})&= \E \left[ \log \left( c_k U_{(k)}^{k-1}  (1-U_{(k)})^{n-k} \right) \right ]  \\
&=\log(c_k) + (k-1)  \E [ \log (  U_{(k)} )]  
\\& \quad \quad +(n-k) \E [ \log ( 1-U_{(k)})].
\label{eq:first_step}
\end{split}
\end{equation}
Now, consider the two expectations on the right side of~\eqref{eq:first_step}, namely $\E[\log( U_{(k)})]$ and $\E[ \log ( 1-U_{(k)})]$. Notice that with $U_{(k)} \sim \mathsf{Beta}(k, n+1-k)$, we also have $1-U_{(k)} \sim \mathsf{Beta}(n+1-k, k)$. 
Therefore, (see, e.g.,~\cite{arnold1992first}),
\begin{align*}
\E [ \log  (U_{(k)} ) ] &= \psi(k ) - \psi(n+1),  \\
\E[ \log  (1-U_{(k)} ) ] &=     \psi( n +1-k) - \psi(n+1),
\end{align*}
where $\psi(\cdot)$ is the digamma function.
Next, we leverage the relationship between the digamma function and the harmonic number, $H_r$ defined in~\eqref{eq:harmonic},
namely
$\psi(k) = H_{k-1} - \gamma,$
where $\gamma$ is the Euler-Mascheroni constant. Putting this all together in~\eqref{eq:first_step}, we have that
\begin{equation}
\begin{split}
&-h(U_{(k)})  =\log(c_k) + (k-1)(\psi(k ) - \psi(n+1)) \notag\\
& \qquad\qquad\qquad +(n-k) ( \psi( n +1-k) - \psi(n+1)) \\
& =\log(c_k) + (k-1)(H_{k-1} - H_{n}) +(n-k) ( H_{n-k} - H_{n}) \\
& =\log(c_k) + (k-1)H_{k-1}  + (n-k) H_{n-k} - (n-1) H_{n}.
\label{eq:second_step}
\end{split}
\end{equation}
Finally, using the fact that $c_k=\frac{n!}{ (k-1)!  (n-k)!}$, and the definition of $T_j = \log(j!) - j H_j$ in~\eqref{eq:harmonic}, we have that
\begin{align}
-h(U_{(k)})& =\log(n!) - \log((k-1)!  ) - \log((n-k)!  ) \notag \\
&\quad + (k-1)H_{k-1}   + (n-k) H_{n-k} - (n-1) H_{n} \notag \\
&= T_n - T_{k-1}  - T_{n-k} + H_{n}.
\label{eq:third_step}
\end{align}
This shows~\eqref{eq:entrUk}. 
To show~\eqref{eq:approximation_of_h(u)}, we let $k<n$ and we  approximate the expression in~\eqref{eq:third_step} by leveraging the following expressions for the series expansion of the harmonic number and the log factorial~\cite{abramowitz1970handbook}, 
\begin{align}
H_k & = \log(k)\!+\!\gamma\!+\!\frac{1}{2k}\!-\!\frac{1}{12k^2}\!+\!\frac{1}{120 k^4}\!-\!   O\left(  \frac{1}{k^6}  \right),\label{eq:Harmonic_number_Expansion}
\\
\log(k!) &= k \log(k) - k + \frac{1}{2} \log ( 2 \pi k) + \frac{1}{12k} \nonumber
\\& \hspace{2.9cm} \qquad -   \frac{1}{360k^3} +  O \left( \frac{1}{k^5}\right). \label{eq:log_k_expansion} 
\end{align} 
Combining~\eqref{eq:Harmonic_number_Expansion} and~\eqref{eq:log_k_expansion}, we arrive at 
\begin{equation}
T_k\!=\! \frac{1}{2} \log \left( \frac{2 \pi k}{\eu} \right) \!-\!(1+\gamma) k \!+\!\frac{1}{6k} -\frac{1}{90k^3}\!+\!O \left( \frac{1}{k^5} \right). \label{eq:Expression_for_Tk}
\end{equation}
%
%
Now, using the expression of $T_k$ in~\eqref{eq:Expression_for_Tk} inside~\eqref{eq:third_step}, and collecting similar terms we find
\begin{align*}
 &h(U_{(k)})  = T_{k-1} +T_{n-k} -T_n-  H_n \\
 &= \frac{1}{2} \log \left( \frac{2 \pi \eu(k-1) (n-k)}{ n^3 } \right)  + \frac{1}{6} \left( \frac{1}{ k-1}+\frac{1}{ n-k}-\frac{4}{n}  \right) \\
 &  - \frac{1}{90}\left(\frac{1}{(k-1)^3} +\frac{1}{(n-k)^3} - \frac{1}{n^3} \right)  +\frac{1}{12n^2} \left(1 -\frac{1}{10 n^2} \right)\\
 & +O \left( \frac{1}{(k-1)^5} \right) +O \left( \frac{1}{(n-k)^5} \right) -O \left( \frac{1}{n^5} \right)+  O\left(  \frac{1}{n^6}  \right).
\end{align*} 
Now,  letting $k=np$ and keeping only the dominant terms in the expression above, we arrive at 
\begin{equation}
\begin{split}
h(U_{(np)}) 
 &= \frac{1}{2} \log \left( \frac{2 \pi \eu \left(p-\frac{1}{n}\right) (1-p)}{ n } \right) +\frac{1}{12n^2} \notag\\
 &\quad + \frac{1}{6n} \left( \frac{1}{ p-\frac{1}{n}}+\frac{1}{1-p}-4  \right) 
+O \left( \frac{1}{n^3} \right).
\end{split}
\end{equation} 
This concludes the proof of~\eqref{eq:approximation_of_h(u)} and also of Lemma~\ref{lem:uniform_entropy}.

\section{Proof of Lemma~\ref{lem:term_T2}}
\label{sec:Limit2}
Recall that by assumption, $\frac{p}{n+1}<\epsilon<p$. Now, define the event 
\begin{equation}
\label{eq:Adef}
\mathcal{A}= \{   p- \epsilon \le  U_{(np)} \le p+ \epsilon \},
\end{equation}
and let $1_{  \mathcal{A}}  $ be the indicator function of the event $\mathcal{A}$ and $1_{  \mathcal{A}^c} $ be the indicator function of the complementary event.
 Now, using the law of total expectation we can write
\begin{align}
&\E \left [ \left( F^{-1} (U_{(np)}) -  F^{-1} (p)   \right)^2 \right] \notag\\
&\quad = \E \left [ \left( F^{-1} (U_{(np)}) -  F^{-1} (p)   \right)^2 1_{  \mathcal{A}^c}   \right]  \notag\\
& \qquad \qquad+  \E \left [ \left( F^{-1} (U_{(np)}) -  F^{-1} (p)   \right)^2 1_{  \mathcal{A}}   \right]. \label{eq:Decomp_bad_good}
\end{align}
We will now analyze the two expectations on the right side of~\eqref{eq:Decomp_bad_good} separately. 

\noindent 
\textbf{{\em First expectation in~\eqref{eq:Decomp_bad_good}.}}
We obtain
\begin{align}
&\E \left [ \left( F^{-1} (U_{(np)}) -  F^{-1} (p)   \right)^2 1_{  \mathcal{A}^c}   \right] \notag\\
 &\stackrel{{\rm{(a)}}}{\le} 2 \E \left [ \left( F^{-1} (U_{(np)})\right)^2 1_{  \mathcal{A}^c} \right] + 2\left( F^{-1} (p)   \right)^2  \mathbb{P}(\mathcal{A}^c)  \notag   \\
&\stackrel{{\rm{(b)}}}{\le}    2     \sqrt{ \E \left[ (X_{(np)})^4 \right] \mathbb{P}(\mathcal{A}^c)  } +  2 \left(F^{-1} (p) \right)^2 \mathbb{P}(\mathcal{A}^c) \notag\\
&\stackrel{{\rm{(c)}}}{\le}   2 \left(      \sqrt{\E \left[ (X_{(np)})^4 \right]   } +   \left(F^{-1} (p) \right)^2 \right)   \sqrt{\mathbb{P}(\mathcal{A}^c) } \notag \\
&\stackrel{{\rm{(d)}}}{\le}   \! 4 \! \left(      \sqrt{ \E \left[ (X_{(np)})^4 \right]   } \!+\!   \left(F^{-1} (p) \right)^2 \right)    \eu^{  - (n+2)  \left(\epsilon -\frac{p}{n+1} \right)^2 },
\label{eq:Bound_on_the_bad_part_int}
\end{align} 
where the labeled inequalities follow from: $\rm{(a)}$ the fact that 
$(a-b)^2 \le 2 (a^2+b^2)$;
$\rm{(b)}$ the Cauchy-Schwarz inequality and
the fact $X_{(np)} \overset{d}{=} F^{-1} (U_{(np)})$; $\rm{(c)}$ the fact that $x \le \sqrt{x}$ for $x \in (0,1)$; and $\rm{(d)}$ Lemma~\ref{lem:UBPAC} below, which is proved in Appendix~\ref{app:UBPAC}.
\begin{lem} \label{lem:UBPAC}
Let $\mathcal{A}= \{   p- \epsilon \le  U_{(np)} \le p+ \epsilon \}$ for $p \in (0,1)$ and $\frac{p}{n+1}<\epsilon<p$. Then,
\begin{align*}
\mathbb{P}(\mathcal{A}^c)& \leq  2 \exp \left( -2 (n+2)  \left(\epsilon -\frac{p}{n+1} \right)^2 \right).
\end{align*}
\end{lem}

\noindent 
\textbf{{\em Second expectation in~\eqref{eq:Decomp_bad_good}.}}
From the definition of $\mathcal{A}$ in~\eqref{eq:Adef}, we know that $U_{(np)} \in [ p- \epsilon,  p+ \epsilon]$; hence, using the Taylor remainder theorem, for any $t \in [ p- \epsilon,  p+ \epsilon]$, where $\tilde{u}$ is some value  between $p$ and $t$, we have
\begin{align}
\label{eq:Taylor1}  
  F^{-1} (t)   
 &= F^{-1} (p) + \frac{{\rm{d}}}{{\rm{d}}u} F^{-1} (u) |_{u=p} \left (t-p \right) \notag\\
 & \quad +  \frac{{\rm{d}}^2}{{\rm{d}}u^2} F^{-1}(u) |_{u=\tilde{u}}  \frac{\left (t- p \right)^2}{2} \notag \\
  &= F^{-1} (p) \!+\! g^{(1)} (p) \left (t\!-\!p \right) \!+\!  \frac{1}{2}  g^{(2)} (\tilde{u}) \left (t\!-\! p \right)^2,
\end{align}
where in the last equality we let 
$ g^{(1)} (p) :=  \frac{{\rm{d}}}{{\rm{d}}u} F^{-1} (u) |_{u=p}$ 
and $ g^{(2)} (\tilde{u}) :=   \frac{{\rm{d}}^2}{{\rm{d}}u^2} F^{-1} (u) |_{u=\tilde{u}}.$

 We next provide an upper bound for $g^{(2)} (\tilde{u})$. Define, 
        \begin{equation}
        C_{p, \epsilon} =\max_{ t\in [ p-\epsilon, p+\epsilon]}  | g^{(2)} (t) | < \infty,
        \end{equation}
        where the last inequality follows since
        \begin{align*}
        g^{(2)} (t) =   \left.  \frac{{\rm{d}}^2}{{\rm{d}}u^2} F^{-1} (u) \right  |_{u=t} &= \left. \frac{ {\rm d}}{ {\rm d} u}   \frac{1}{   f (F^{-1}(u))  }\right  |_{u=t} 
        \\& = -   \frac{  f'(F^{-1}(t))  }{  ( f (F^{-1}(t))  )^3 }.
        \end{align*}
Therefore, for sufficiently small $\epsilon>0$, we have that $C_{p,\epsilon}<\infty$  if  $f (F^{-1}(p)) \neq 0$,    and    $f'(F^{-1}(t))$ is continuous at $t=p$; these two conditions are guaranteed by the assumptions of Lemma~\ref{lem:term_T2}. 
Thus, plugging this into \eqref{eq:Taylor1},
\begin{align}
\label{eq:Taylor2}  
\left(F^{-1} (t)   - F^{-1} (p) \right)^2  &\leq \left(g^{(1)} (p)\right)^2 \left (t-p \right)^2  +  \frac{C_{p, \epsilon}^2 }{4}   \left (t- p \right)^4 \notag \\
 & \qquad+ C_{p, \epsilon} \left \lvert g^{(1)} (p) \left (t-p \right)^3\right \lvert.
\end{align}
Now using \eqref{eq:Taylor2},
\begin{align}
  &   \E \left [ \left( F^{-1} (U_{(np)}) -  F^{-1} (p)   \right)^2 1_{  \mathcal{A}}   \right] \notag  \\
    &\leq \left(g^{(1)} (p) \right)^2   \E \left [ ( U_{(np)} -p  )^2    1_{  \mathcal{A}}   \right] +\frac{C_{p, \epsilon}^2}{4}\E  [ (U_{(np)} - p)^4     1_{  \mathcal{A}}  ] \notag
\\& \quad +  C_{p, \epsilon}   \E \left [   \left \lvert g^{(1)} (p)   (U_{(np)} - p)^3 1_{  \mathcal{A}} \right\lvert  \right].\label{eq:After_Taylor} 
\end{align}
Next, we individually upper bound each term on the right side of~\eqref{eq:After_Taylor}. 

\noindent 
\textbf{{\em First term in~\eqref{eq:After_Taylor}.}} 
We have
\begin{equation}
\begin{split}
     \E \left [ \left( U_{(np)}  -p \right)^2    1_{  \mathcal{A}}   \right]  
   & \leq \E \left [ \left( U_{(np)} -p  \right)^2       \right]  \\
   & = {\rm Var}(U_{(np)})+ \frac{p^2}{ (n+1)^2},\label{eq:Bound_first_term}
\end{split}
\end{equation} 
where the equality follows since 
\begin{align*}
\E [( U_{(np)} -p  )^2 ] &=  \E [( U_{(np)} - \E[ U_{(np)}] + \E[ U_{(np)}] -p   )^2 ] \notag\\
&=  {\rm Var}(U_{(np)})+  \left(\E[ U_{(np)}] -p   \right)^2, 
\end{align*}
where $\E[ U_{(np)}]=\frac{np}{n+1}$, so $\left (\E[ U_{(np)}] - p   \right )^2  =  (\frac{p}{n+1})^2$. From~\eqref{eq:Bound_first_term}, the first term on the right side of~\eqref{eq:After_Taylor} can be upper bounded as 
\begin{align}
&  \left(g^{(1)} (p) \right)^2    \E \left [ ( U_{(np)} -p  )^2    1_{  \mathcal{A}}   \right]  \notag\\
    &\quad \leq \left(g^{(1)} (p) \right)^2  \left(  {\rm Var}(U_{(np)})+ \frac{p^2}{ (n+1)^2} \right).\label{eq:Bound_first_term_full}
\end{align} 
%

\noindent 
\textbf{{\em Second term in~\eqref{eq:After_Taylor}.}}      
We obtain
\begin{align}
    &  \frac{C_{p,\epsilon}^2}{4} \E \left [ \left (U_{(np)} -p \right)^4     1_{  \mathcal{A}}   \right]  \notag
       \\& \leq   \frac{C_{p,\epsilon}^2}{4}   \E \left [  \left ( U_{(np)}  -\frac{pn}{n+1} + \frac{pn}{n+1} - p \right)^4       \right] \notag
\\&       \stackrel{{\rm{(a)}}}{\leq} 2 C_{p,\epsilon}^2  \left (  \left (\frac{pn}{n+1} -p \right)^4  +  \E \left [\left( U_{(np)} - \frac{pn}{n+1} \right)^4       \right] \right) \notag
\\&=  2 C_{p,\epsilon}^2   \left (   \frac{p^4}{(n+1)^4}   +  \E \left [\left(  U_{(np)}  - \frac{pn}{n+1} \right)^4       \right] \right) \notag
       \\& \stackrel{{\rm{(b)}}}{=}  2 C_{p,\epsilon}^2 \left (   \frac{p^4}{(n+1)^4}   +  \Theta\left( \frac{1}{n^2} \right)  \right),
        \label{eq:Bounding_Second_term_in_square-diff0}       
\end{align}
where the labeled (in)equalities follow from: 
${\rm{(a)}}$ the fact that $(a+b)^4 \le 8 (a^4+b^4)$, and $\rm{(b)}$
 the fact that if $W \sim \mathsf{Beta}(\alpha,\beta)$, then (see, for example~\cite{papoulis1962fourier})
\begin{align}
&\E[ (W-\E[W])^4 ] \notag\\
&\hspace{.3cm} \!=\! \left( \frac{\alpha}{\alpha\!+\!\beta} \right)^4    \cdot 
\frac{3 (\alpha^2 \beta^2 \!+\! 2 \alpha^2 \beta \!+\! \alpha \beta^3 \!-\! 2 \alpha \beta^2 \!+\! 2 \beta^3)} {\alpha^3 (\alpha \!+\! \beta \!+\! 1) (\alpha \!+\! \beta \!+\! 2) (\alpha \!+\! \beta \!+\! 3)};
\label{eq:beta_res0}
\end{align} 
and, in particular, for $U_{(np)} \sim \mathsf{Beta}(np, n+1-np)$, 
by 
setting $\alpha=pn$ and $\beta= n-pn+1$ in~\eqref{eq:beta_res0} 
\begin{equation*}
\E\left [\left ( U_{(np)}  -  \frac{pn}{n+1}   \right)^4  \right] =  \Theta\left ( \frac{1}{n^2} \right).
\end{equation*}

\textbf{{\em Third term in~\eqref{eq:After_Taylor}.}}
This term can be upper bounded using Jensen's inequality (as $\mathbb{E}[|Y|^3] = \mathbb{E}[(|Y|^4)^{3/4}] \leq (\mathbb{E}[|Y|^4])^{3/4}$) and the same steps as we used to get the bound in~\eqref{eq:Bounding_Second_term_in_square-diff0}. Indeed,
       \begin{align}
&  C_{p, \epsilon}   \E \left [   \left \lvert g^{(1)} (p)   (U_{(np)} - p)^3 1_{  \mathcal{A}} \right\lvert  \right] \notag \\
& \qquad  \le   C_{p,\epsilon}   \left \lvert g^{(1)} (p)   \right \lvert \E \left [    \left | U_{(np)} -p \right|^3 1_{  \mathcal{A}}   \right] \notag\\
&\qquad \le   C_{p,\epsilon} \left \lvert g^{(1)} (p)   \right \lvert  \left(  \E \left [    \left | U_{(np)} - p \right|^4   \right]  \right )^{  \frac{3}{4} } \notag\\
&\qquad \leq 8^{3/4} C_{p,\epsilon} \left \lvert g^{(1)} (p)   \right \lvert  \left (   \frac{p^3}{(n+1)^3}   +  \Theta \left( \frac{1}{n^\frac{3}{2}}\right) \right) . \label{eq:bound_on_third_term}
\end{align} 
Collecting~\eqref{eq:After_Taylor} together with the bounds in~\eqref{eq:Bound_first_term_full},~\eqref{eq:Bounding_Second_term_in_square-diff0}, and~\eqref{eq:bound_on_third_term},
 \begin{align}
  & \E \left [ \left( F^{-1} (U_{(np)}) -  F^{-1} (p)   \right)^2 1_{  \mathcal{A}}   \right] \notag\\
    & \qquad  \qquad \leq \left(g^{(1)} (p) \right)^2  {\rm Var}(U_{(np)})+  C_{p,\epsilon}    O \left( \frac{1}{n^\frac{3}{2}} \right). \label{eq:Final_bound_on_good_term}
 \end{align}
Therefore, plugging~\eqref{eq:Bound_on_the_bad_part_int} and~\eqref{eq:Final_bound_on_good_term} in~\eqref{eq:Decomp_bad_good},
\begin{align*}
&\E \left [ \left( F^{-1} (U_{(np)}) -  F^{-1} (p)   \right)^2 \right] \notag\\
  &\quad \leq  4  \left(      \sqrt{ \E \left[ (X_{(np)})^4 \right]   } +   \left(F^{-1} (p) \right)^2 \right)    \eu^{  - (n+2)  \left(\epsilon -\frac{p}{n+1} \right)^2 } \notag\\
  &\qquad \qquad + \left(g^{(1)} (p) \right)^2  {\rm Var}(U_{(np)})+ C_{p,\epsilon}  O \left( \frac{1}{n^\frac{3}{2}} \right) \notag \\
& \quad =4       \sqrt{ \E \left[ (X_{(np)})^4 \right]   }    \eu^{  - (n+2)  \left(\epsilon -\frac{p}{n+1} \right)^2 } \notag\\
  &\qquad \qquad + \frac{  {\rm Var}(U_{(np)})}{ \left(  f (F^{-1}(p))  \right)^{2} }+ C_{p,\epsilon}  O \left( \frac{1}{n^\frac{3}{2}} \right) ,
\end{align*} 
where the last equality follows by absorbing non-dominant terms inside the big-$O$ and recalling that 
$
 g^{(1)} (p) =    \left[  f (F^{-1}(p))  \right]^{-1}. 
$
This concludes the proof of Lemma~\ref{lem:term_T2}.


%


 \section{Proof of Lemma~\ref{lem:UBPAC}} 
\label{app:UBPAC}
Using the definition of $\mathcal{A}$ in~\eqref{eq:Adef},
\begin{align}
\label{eq:first_boundLem7}
\mathbb{P}(\mathcal{A}^c)&=\mathbb{P} [  U_{(np)} \le p-\epsilon ] +\mathbb{P} \left [  U_{(n p)} \ge  p+\epsilon \right ] \nonumber \\
&= \!\mathbb{P} [  U_{(n (1-p)+1)} \!\ge\!  1\!-\!p\!+\!\epsilon ]\!+\!\mathbb{P} \left [  U_{(n p)} \!\ge\!  p\!+\!\epsilon \right ],
\end{align} 
where in the last equality we  have used the fact that $U_{(np)} \stackrel{d}{=} 1-U_{(n (1-p)+1)}$  where $U_{(n (1-p)+1)} \sim \mathsf{Beta}(n+1-np, np)$ since $U_{(np)} \sim \mathsf{Beta}(np, n+1-np)$. 

To upper bound $\mathbb{P}(\mathcal{A}^c)$ in~\eqref{eq:first_boundLem7}, we leverage the fact that Beta-distributed random variables $U_{(np)}$ and $U_{(n (1-p)+1)}$ are $\sigma_0$-sub-Gaussian with $\sigma_0 <\frac{1}{ 4(\alpha + \beta+1)} = \frac{1}{ 4(n+2)}$ (see \cite[Thm.~1]{marchal2017sub}). Recall (see, for example, \cite{boucheron2013concentration}), that a random variable $X$ is sub-Gaussian with parameter $\sigma_0$ if for every $\lambda \in \mathbb{R}$,
\begin{equation}
\label{eq:subG1}
\mathbb{E}[e^{\lambda (X - \mathbb{E}[X])}] \leq e^{\lambda^2 \sigma_0 /2}.
\end{equation}
Moreover, if $X$ is $\sigma_0$-sub-Gaussian, then for all $t \geq 0$,
\begin{equation}
\label{eq:subG2}
\mathbb{P}(X - \mathbb{E}[X] \geq t) \leq  e^{-\frac{t^2}{2 \sigma_0}}.
\end{equation}
Now, let us consider the first term on the right side of~\eqref{eq:first_boundLem7}. Using that $ \mathbb{E}[U_{(n (1-p)+1)}]  = \frac{n+1-np}{n+1}$, 
 \begin{align}
  \mathbb{P} &[  U_{(n (1-p)+1)} \ge  1-p+\epsilon ]  \notag\\
    &=\mathbb{P} \left [  U_{(n (1-p)+1)}   - \mathbb{E}[U_{(n (1-p)+1)}]   \ge   \epsilon -\frac{p}{n+1} \right] \notag \\
      & \stackrel{{\rm{(a)}}}{\le}    \exp \left( -\frac{1}{2 \sigma_0} \left(  \epsilon-\frac{p}{n+1}  \right)^2 \right) \notag
\\& \stackrel{{\rm{(b)}}}{\le}    \exp \left( -2 (n+2)  \left( \epsilon -\frac{p}{n+1}  \right)^2 \right),
  \label{eq:cheby1}
  \end{align}
where $\rm{(a)}$ follows from the bound in~\eqref{eq:subG2}, and $\rm{(b)}$ is due to the fact that $\sigma_0 <\frac{1}{ 4(n+2)}$.

Similarly, for the second term on the right side of~\eqref{eq:first_boundLem7}, using that $\mathbb{E}[U_{(np)}] = \frac{np}{n+1}$, we obtain 
 \begin{align}
\mathbb{P} \left [  U_{(n p)} \ge  p+\epsilon \right ] 
&=\mathbb{P} \left [  U_{(np)}   - \mathbb{E}[U_{(np)}]   \ge \frac{p}{n+1}  +\epsilon \right] \notag \\
& \stackrel{{\rm{(a)}}}{\le}    \exp \left( -2 (n+2)  \left( \frac{p}{n+1}  +\epsilon \right)^2 \right) \notag
\\& \leq \exp \left( -2 (n+2) \epsilon^2\right ),
\label{eq:cheby2}
  \end{align}
where, again, $\rm{(a)}$ follows from the bound in~\eqref{eq:subG2}, and  the fact that $\sigma_0 <\frac{1}{ 4(n+2)}$.

Now, by summing~\eqref{eq:cheby1} and~\eqref{eq:cheby2}, we sind that $\mathbb{P}(\mathcal{A}^c)$ in~\eqref{eq:first_boundLem7} can be upper bounded as
\begin{align*}
&\mathbb{P}(\mathcal{A}^c) \notag\\
& \leq \exp \left( -2 (n+2)  \left(   \epsilon -\frac{p}{n+1} \right)^2 \right) + \exp \left( -2 (n+2) \epsilon^2\right )
\\& \leq 2 \exp \left( -2 (n+2)  \left(  \epsilon-\frac{p}{n+1} \right)^2 \right),
\end{align*} 
concluding the proof of Lemma~\ref{lem:UBPAC}.


\section{Proof of Lemma~\ref{lem:BOund_On_Moments}}
\label{app:BOund_On_Moments}
Recall from~\eqref{eq:OrdStatUn} that $X_{(k)} \overset{d}{=} F^{-1}(U_{(k)})$, where $U_{(k)}$ is the order statistic of a sample of uniform random variables. Thus, 
\begin{equation}
\begin{split}
&\E [|  X_{(k)}|^q ] = \E \left [|  F^{-1}(U_{(k)})|^q  \right]\\
&\leq    \left( \E [|X|^r] \right)^{\frac{q}{r}}  \E \left[    \left(   \frac{1}{ \min (U_{(k)},1-U_{(k)}) }  \right)^{\frac{q}{r}}     \right],  \label{eq:lembound1}
   \end{split}
   \end{equation} 
where the  inequality follows from Lemma~\ref{lem:Bound_on_the_quantile_function} given below and proved in Appendix~\ref{app:AuxLemma}.
\begin{lem} \label{lem:Bound_on_the_quantile_function}   
Let $F$ be the cdf of the random variable $X$. Then, for any $r >0$,
\begin{align}
|F^{-1}(u)| \le \left( \frac{\E [|X|^r] }{ \min (u,1-u) }\right)^{  \frac{1}{r}}, \, \qquad u \in (0,1).
\end{align}
\end{lem}
Next, we upper bound the right side of \eqref{eq:lembound1} using the following:
\begin{align}
 \E \left[    \left(   \frac{1}{ \min (U_{(k)},1-U_{(k)}) }  \right)^{\frac{q}{r}}     \right]  & \stackrel{{\rm{(a)}}}{\le} \E \left[   \left(  \frac{1}{  U_{(k)} ( 1-U_{(k)}) }  \right)^{\frac{q}{r}}  \right] \nonumber
\\ & \stackrel{{\rm{(b)}}}{\le}   C_{n,k,q,r} .  \label{eq:lembound2}
   \end{align} 
In the above, the inequalities follow from the facts that  $\rm{(a)}$ $\min\{a,b\}\geq \frac{ab}{a+b}$
and $\rm{(b)}$ $U_{(k)} \sim \mathsf{Beta}(k, n+1-k)$; hence,
    \begin{align*}
 &\E \left[   \left(  \frac{1}{  U_{(k)} ( 1-U_{(k)}) }  \right)^{\frac{q}{r}}  \right] \notag\\
 &\quad =    \frac{\Gamma(n+1)}{ \Gamma(k)  \Gamma(n+1-k)}  \int_0^1   t^{k-1-\frac{q}{r}}
 (1-t)^{n-k-\frac{q}{r}}   {\rm d} t\\
 &\quad = \frac{\Gamma(n+1)}{ \Gamma(k)  \Gamma(n+1-k)}
 \\&  \quad \qquad \cdot \begin{cases}
 \frac{\Gamma\left(k-\frac{q}{r} \right) \Gamma\left(n-k-\frac{q}{r}+1 \right)  }{\Gamma \left(n-\frac{2q}{r}+1 \right) } &    k> \frac{q}{r} \text{ and } n+1 -k >\frac{q}{r}\\
 \infty & \text{ else} 
 \end{cases}  \\
 &\quad =C_{n,k,q,r}.
  \end{align*} 
  Plugging \eqref{eq:lembound2} into \eqref{eq:lembound1}, we have shown
  \[\E [|  X_{(k)}|^q ]  \leq   C_{n,k,q,r}   \left( \E [|X|^r] \right)^{\frac{q}{r}},  \]
  as desired.
  
Finally,  we notice that if $q$ and $r$ are fixed with $k = np$ for fixed $p$, then as $n \rightarrow \infty$, we have $ k = np > \frac{q}{r}$ and $n+1 -k = n(1-p)+ 1 >\frac{q}{r}$, eventually. Then,
by~\cite[eq.~6.1.46]{abramowitz1970handbook}, for $ a,b\in \mathbb{R}$,
\begin{align*}
\lim_{x \to \infty} \frac{\Gamma(x +a) }{\Gamma(x +b) \,\, x^{b-a}}=1,
\end{align*}
which leads to the conclusion that
\begin{align*}
& \lim_{n \to \infty}  C_{n,pn,q,r} \\
 &=  \lim_{n \to \infty}   \frac{\Gamma(n+1)}{\Gamma \left(n-\frac{2q}{r}+1 \right) }   \frac{ \Gamma\left(pn-\frac{q}{r} \right) \ }{\Gamma(pn)}  \frac{ \Gamma\left(n(1-p)-\frac{q}{r}+1 \right)}
 {  \Gamma\left(n(1-p)+1 \right) }  \\
 &=\lim_{n \to \infty} n^{-\frac{2q}{r}} (pn)^{\frac{q}{r}} (n(1-p))^{\frac{q}{r}} = \left( p (1-p) \right)^{  \frac{q}{r} }.
\end{align*} 
This concludes the proof of Lemma~\ref{lem:BOund_On_Moments}.




\section{Proof of Lemma~\ref{lem:Bound_on_the_quantile_function}} 
\label{app:AuxLemma}


First, recall that for a uniform random variable over $[0,1]$, denoted as $U$, we have $X \overset{d}{=} F^{-1}(U)$. Therefore,
\begin{align*}
\E[|X|^r] &= \E[|F^{-1}(U)|^r] = \int_0^1 |F^{-1}(t)|^r \ {\rm d} t.
\end{align*}

First, assume  that $0 \le F(0)  \le  u\leq 1$. Then,
\begin{align}
 \int_0^1 \! |F^{-1}(t)|^r {\rm d} t  \!\ge\!  \int_u ^1 \! |F^{-1}(t)|^r {\rm d} t  \!\ge\!  (1-u)  |F^{-1}(u)|^r,\label{eq:Bound_1} 
\end{align} 
where in the last inequality we have used that  $F^{-1}(t)$ is positive for $t \geq u \geq F(0)$, implying $|F^{-1}(u)|^r \leq |F^{-1}(t)|^r$ for $t \in [u,1]$, and that $F^{-1}(x)$ is non-decreasing in $x$; hence, $F^{-1}(u) \leq F^{-1}(t)$ for $t \in [u,1]$. 



Now assume, on the other hand, $0 \le  u\le F(0) \leq 1$. Then, 
\begin{align}
 \int_0^1 |F^{-1}(t)|^r {\rm d} t  &\ge  \int_0 ^u |F^{-1}(t)|^r {\rm d} t  \ge  u  |F^{-1}(u)|^r. \label{eq:Bound_2} 
\end{align} 
In the last inequality, we have used  that $F^{-1}(t)$ is negative for $t \leq u\le F(0)$,  implying that $|F^{-1}(u)| \leq |F^{-1}(t)|$ and that $F^{-1}(x)$ is non-decreasing in $x$; hence, $F^{-1}(t) \leq F^{-1}(u)$ for $t \in [0,u]$. Thus,
$|F^{-1}(u)|^r \leq |F^{-1}(t)|^r$.

The bounds in~\eqref{eq:Bound_1} and~\eqref{eq:Bound_2} imply that 
\begin{align*}
|F^{-1}(u)| \le  \left(\frac{\E [|X|^r] }{ u }\right)^{\frac{1}{r}}, \, \qquad \text{ for }  u \in [0,F(0)],\\
|F^{-1}(u)| \le  \left(\frac{\E [|X|^r] }{ 1-u }\right)^{\frac{1}{r}}, \, \qquad \text{ for } u \in [F(0), 1].
\end{align*}
Taking the largest of the two bounds concludes the proof of Lemma~\ref{lem:Bound_on_the_quantile_function}.


\section{ Proof  of Lemma~\ref{lem:Log_Term_bound} }
\label{app:lem:Log_Term_bound}
Choose some $q \in [1,\infty]$ and define an event $\mathcal{A}\!=\! \{   p- \epsilon \!\le\!  U_{(np)} \le p\!+\! \epsilon \}$, where we assume that
\begin{align*}
p>\epsilon> \max  \left\{  \frac{p}{n+1},  \frac{|(q-2) p-q +1|}{q(n-1)+2}  \right \} .
\end{align*} 
Next, notice that we can write
\begin{align}
&\E \left[ \log (f(F^{-1} (U_{(np)} )))  \right] = \E \left[ \log ( f ( F^{-1} (U_{(np)}))))   1_{  \mathcal{A}}  \right]  \notag\\
& \qquad \qquad \qquad + \E \left[ \log ( f ( F^{-1} (U_{(np)} )))  1_{  \mathcal{A}^c}  \right]   .\label{eq:Split_Sub_Cases}
\end{align} 
We now analyze each expectation on the right side of~\eqref{eq:Split_Sub_Cases} separately. 

\noindent \textbf{{\em First expectation in~\eqref{eq:Split_Sub_Cases}.}} Using Taylor's remainder theorem we have that
\begin{align*}
 &\log ( f ( F^{-1} (u ))) = \log ( f ( F^{-1} (p)))+\frac{f' ( F^{-1} ( \tilde{u} )) }{ (f ( F^{-1} ( \tilde{u} )))^2 }  (u -p),  
\end{align*} 
where $\tilde{u}$ is some number between $p$ and $u$. Therefore, 
using the definition
\begin{equation*}
C^{(2)}_\epsilon= \max_{ p-\epsilon \le u \le p+\epsilon}  \left| \frac{f' \left( F^{-1} \left(u \right) \right) }{ \left( f \left( F^{-1} \left( u\right)  \right) \right)^2 }  \right|,
\end{equation*} 
we find that
\begin{align}
\label{eq:FEPA1}
 &\E \left[ \log ( f ( F^{-1} (U_{(np)} )))   1_{  \mathcal{A}}  \right] \notag\\
  &\leq  \log ( f ( F^{-1}(p))) \mathbb{P} \left[ \mathcal{A} \right]+ C^{(2)}_\epsilon  \E \left[  |U_{(np)} -p|  1_{  \mathcal{A}}  \right].
\end{align} 
Next, we notice that
\begin{equation}
\label{eq:FEPA2}
 \E \left[  |U_{(np)} -p|  1_{  \mathcal{A}}  \right] \!\stackrel{{\rm{(a)}}}{\le}\!  \sqrt{ \E \left[  |U_{(np)} -p|^2   \right] } \!\stackrel{{\rm{(b)}}}{=}  \! O ({1}/{\sqrt{n} }),
\end{equation} 
where in the above, 
$\rm{(a)}$  follows using Cauchy-Schwarz inequality and the fact that $\mathbb{P} \left[ \mathcal{A} \right] \leq 1$ and $\rm{(b)}$ using Remark~\ref{rem:Uni}.
Combining \eqref{eq:FEPA1} and \eqref{eq:FEPA2} we have the bound
\begin{align}
\label{eq:FEPA}
 &\E \left[ \log ( f ( F^{-1} (U_{(np)} )))   1_{  \mathcal{A}}  \right] \notag \\
        &\qquad \leq  \log ( f ( F^{-1}(p))) \mathbb{P} \left[ \mathcal{A} \right]+ C^{(2)}_\epsilon O \left( \frac{1}{\sqrt{n} } \right).
\end{align}

\noindent \textbf{{\em Second expectation in~\eqref{eq:Split_Sub_Cases}.}}
First recall that $U_{(np)} \sim \mathsf{Beta}(np, n+1-np)$ and let $f_{\alpha, \beta}$ denote the beta distribution with parameters $\alpha := np$ and $\beta := n+1-np$. Then, using the bound $\log(x) \le x$, we have
\begin{align}
&\E \left[ \log( f ( F^{-1} (U_{(np)}) ) )  1_{  \mathcal{A}^c}  \right]  \leq  \E \left[  f ( F^{-1} (U_{(np)} ) )  1_{  \mathcal{A}^c}  \right] \notag \\
&\qquad =  \int_{0}^1 1_{  \mathcal{A}^c} f_{\alpha, \beta }(u)     f ( F^{-1} (u))  {\rm d} u.
 \label{eq:Bound_on_log_of_bad_term_0}
\end{align} 
Next, by H{\" o}lder's inequality,  for any $q , r \in [1,\infty]$ such that $\frac{1}{q}+\frac{1}{r}=1$, we find
\begin{align}
& \int_{0}^1 1_{  \mathcal{A}^c} f_{\alpha, \beta }(u)     f ( F^{-1} (u))  {\rm d} u \notag \\
&\leq \left(  \int_{0}^1 1_{  \mathcal{A}^c} f_{\alpha, \beta }^q(u) {\rm d} u  \right)^{ \frac{1}{q}}  \left( \int_{0}^1    \left( f ( F^{-1} (u)) \right )^{r}  {\rm d} u \right )^{  \frac{1}{r} }.  \label{eq:Bound_on_log_of_bad_term_1}
\end{align}

Next we simplify the two terms on the right side of \eqref{eq:Bound_on_log_of_bad_term_1}. First notice that $f_{\alpha, \beta }^q(u) =  \frac{c_{\alpha^*, \beta^*}}{c_{\alpha,\beta}^q} f_{\alpha^*, \beta^* }(u)$ where 
\begin{equation}
\label{eq:star_defs}
\alpha^*=q (\alpha-1)+1, \quad  \beta^*=q (\beta-1)+1,
\end{equation}
and
\begin{equation*}
c_{i,j} = \frac{\Gamma(i) \Gamma(j)}{\Gamma(i+j)}, \quad i \in \{\alpha, \alpha^*\}, \ j \in \{\beta, \beta^*\}.
\end{equation*}
It follows that,
\begin{align}
 \left(  \int_{0}^1 1_{  \mathcal{A}^c} f_{\alpha, \beta }^q(u) {\rm d} u  \right)^{ \frac{1}{q}}   &= \frac{c_{\alpha^*, \beta^*}^{ \frac{1}{q}} }{c_{\alpha,\beta}} \left(  \int_{0}^1 1_{  \mathcal{A}^c} f_{\alpha^*, \beta^* }(u) {\rm d} u  \right)^{ \frac{1}{q}}      \notag
\\
&= \frac{c_{\alpha^*, \beta^*}^{ \frac{1}{q}} }{c_{\alpha,\beta}} \mathbb{P}^{ \frac{1}{q}} \left( U_{\alpha^*, \beta^*}  \in \mathcal{A}^c \right)   ,  \label{eq:Bound_on_log_of_bad_term_2}
\end{align} 
where in the final equality $U_{\alpha^*,\beta^*}$ denotes a beta random variable with parameters $\alpha^*$ and $\beta^*$.
For the second term on the right side of \eqref{eq:Bound_on_log_of_bad_term_1}, we use a change of variables with $x = F^{-1}(u)$ with $ {\rm d} x = (f(F^{-1}(u)))^{-1}  {\rm d} u$ to get
\begin{align}
 \left( \int_{0}^1    \left( f ( F^{-1} (u)) \right )^{r}  {\rm d} u \right )^{  \frac{1}{r} } &= \left( \int_{-\infty}^\infty  ( f (x))^{ r+1}  {\rm d} x \right )^{ \frac{1}{r}}  \notag \\
&=  \left( \| f \|_{r+1}\right )^{\frac{r+1}{r}}.  \label{eq:Bound_on_log_of_bad_term_3}
\end{align}
Therefore, putting together \eqref{eq:Bound_on_log_of_bad_term_0} -- \eqref{eq:Bound_on_log_of_bad_term_3}, we have shown
\begin{align}
&\E \left[ \log( f ( F^{-1} (U_{(np)}) ) )  1_{  \mathcal{A}^c}  \right]  \notag\\
&\qquad \leq \frac{c_{\alpha^*, \beta^*}^{ \frac{1}{q}} }{c_{\alpha,\beta}} \mathbb{P}^{ \frac{1}{q}} \left( U_{\alpha^*, \beta^*}  \in \mathcal{A}^c \right)   \left( \| f \|_{r+1}\right )^{\frac{r+1}{r}}.  \label{eq:Bound_on_log_of_bad_term}
\end{align} 
We now focus on further upper bounding the right-hand side of~\eqref{eq:Bound_on_log_of_bad_term}. Towards this end, we start by noting that by the definitions in \eqref{eq:star_defs} and the fact that $\alpha := np$ and $\beta := n+1-np$, we find $\alpha^*+\beta^*= q(\alpha+\beta-2) +2=q(n-1) +2$, therefore
\begin{equation*}
\frac{\alpha^*}{\alpha^*+\beta^*}= \frac{q(pn-1) +1}{q(n-1) +2}, \qquad 
\frac{\beta^*}{\alpha^*+\beta^*}= \frac{q(1-p)n +1}{q(n-1) +2}.
\end{equation*}
Moreover, in Appendix~\ref{app:eq:Stirling_bounds_on_c}, it is shown that
\begin{equation}
\label{eq:ConstIntermStep1}
\frac{c_{\alpha^*, \beta^*}^{ \frac{1}{q}} }{c_{\alpha,\beta}} \leq C_q n^{\frac{1}{2}\left(1 -\frac{1}{q}\right)}.
\end{equation} 
where the universal constant $C_q := {\rm{e}}^{1+\frac{2}{q}} \left(\sqrt{2 \pi }\right)^{\frac{1}{q}-1}  q^{- \frac{1}{2q}}$.

Finally, we note that from properties of the beta distribution, namely that $U_{\alpha^*, \beta^*} \stackrel{d}{=} 1-U_{ \beta^*,\alpha^*}$, we find
\begin{align}
& \mathbb{P}(U_{\alpha^*, \beta^*} \in \mathcal{A}^c)=\mathbb{P} [  U_{\alpha^*, \beta^*} \le p-\epsilon ] +\mathbb{P} \left [  U_{\alpha^*, \beta^*} \ge  p+\epsilon \right ] \notag \\
&=\mathbb{P} [  U_{\beta^*, \alpha^*} \ge 1- p + \epsilon ] +\mathbb{P} \left [  U_{\alpha^*, \beta^*} \ge  p+\epsilon \right ],\label{eq:Concentration_inequal_0} 
\end{align} 
Now focus on the first term on the right side of \eqref{eq:Concentration_inequal_0}, we have
\begin{align}
& \mathbb{P} [  U_{\beta^*, \alpha^*} \ge 1- p +\epsilon ]  \notag \\
&=  \mathbb{P}  \left[ U_{ \beta^*,\alpha^*}-\frac{\beta^*}{\alpha^*+\beta^*} \ge  1-p+\epsilon -\frac{\beta^*}{\alpha^*+\beta^*}  \right] \notag\\
&\stackrel{{\rm{(a)}}}{\le}  \mathbb{P} \left [ U_{ \beta^*,\alpha^*}-\frac{\beta^*}{\alpha^*+\beta^*} \ge  \epsilon- \frac{ |(q-2) p-q +1|}{q(n-1)+2}  \right] \notag \\
&\stackrel{{\rm{(b)}}}{\le}  \eu^{- 2(q (n-1)+3) \left( \epsilon- \frac{|(q-2) p-q +1|}{q(n-1)+2}  \right)^2   } \notag \\
& \le   \eu^{- 2(n+2) \left( \epsilon- \frac{|(q-2) p-q +1|}{q(n-1)+2}  \right)^2   } ,\label{eq:Concentration_inequal_1} 
\end{align} 
where the labeled steps follow from:
$\rm{(a)}$
noting that 
\begin{align*}
 1-p+\epsilon -\frac{\beta^*}{\alpha^*+\beta^*} 
 &=\epsilon+ \frac{(q-2) p-q+1}{q(n-1)+2} \notag\\
 & \ge \epsilon- \frac{|(q-2) p-q +1|}{q(n-1)+2}, 
 \end{align*}
and $\rm{(b)}$ applying~\eqref{eq:subG2} since $U_{ \beta^*,\alpha^*}$ is $\sigma_0$-sub-Gaussian random variables with $\sigma_0 <\frac{1}{ 4(\alpha^*+ \beta^*+1)}=\frac{1}{4(q(n-1) +3 ) }$ ~\cite[Thm.~1]{marchal2017sub}. 

Similarly, since
\begin{align*}
p+\epsilon -\frac{\alpha^*}{\alpha^*+\beta^*} 
 &=\epsilon- \frac{(q-2) p-q +1}{q(n-1)+2} \notag\\
 & \ge \epsilon- \frac{|(q-2) p-q+1|}{q(n-1)+2};
 \end{align*}
along with the fact that $U_{\alpha^*, \beta^*}$ is $\sigma_0$-sub-Gaussian again with $\sigma_0 <\frac{1}{ 4(\alpha^*+ \beta^*+1)}=\frac{1}{4(q(n-1) +3 ) }$ , we have the same bound for the second term on the right side of \eqref{eq:Concentration_inequal_0}:
\begin{align}
&\mathbb{P} \left [  U_{\alpha^*, \beta^*} \ge  p+\epsilon \right ] \notag \\
&\le \mathbb{P} \left [  U_{\alpha^*, \beta^*}  -\frac{\alpha^*}{\alpha^*+\beta^*} \ge  \epsilon- \frac{|(q-2) p-q +1|}{q(n-1)+2} \right ] \nonumber \\
& \le   \eu^{- 2(n+2) \left( \epsilon- \frac{|(q-2) p-q+1|}{q(n-1)+2}  \right)^2   }.\label{eq:Concentration_inequal_2} 
\end{align} 
Thus, we have shown in \eqref{eq:Concentration_inequal_0}-\eqref{eq:Concentration_inequal_2}
\begin{align}
& \mathbb{P}(U_{\alpha^*, \beta^*} \in \mathcal{A}^c) \leq  2 \eu^{- 2(n+2) \left( \epsilon- \frac{|(q-2) p-q +1|}{q(n-1)+2}  \right)^2   }.\label{eq:Concentration_inequal} 
\end{align}
Now, combining~\eqref{eq:Bound_on_log_of_bad_term} with~\eqref{eq:ConstIntermStep} and~\eqref{eq:Concentration_inequal} we arrive at
\begin{align}
\label{eq:SEPAC}
&\E \left[ \log \left( f ( F^{-1}(U_{(np)} )) \right)  1_{  \mathcal{A}^c}  \right] \notag\\
& \leq 2 C_q \left( \| f \|_{r+1}\right )^{\frac{r+1}{r}} n^{\frac{1}{2}\left(1 -\frac{1}{q}\right)}  \eu^{- 2(n+2) \left( \epsilon- \frac{|(q-2) p-q +1|}{q(n-1)+2}  \right)^2   }.
\end{align}

Consequently, combining~\eqref{eq:Split_Sub_Cases},~\eqref{eq:FEPA},  and~\eqref{eq:SEPAC}, we obtain 
\begin{align*}
&\E \left[ \log ( f ( F^{-1} (U_{(np)} )))  \right]-  \log ( f( F^{-1} (p)))  \\
& \le  \lvert   \log ( f ( F^{-1} (p )) )  \lvert    \mathbb{P} [ \mathcal{A}^c] +C_\epsilon^{(2)} O \left( \frac{1}{\sqrt{n}} \right) \notag\\
& \quad + 2 C_q \left( \| f \|_{r+1}\right )^{\frac{r+1}{r}}  n^{\frac{1}{2}\left(1 -\frac{1}{q}\right)}  \eu^{- 2(n+2) \left( \epsilon- \frac{|(q-2) p-q +1|}{q(n-1)+2}  \right)^2  } \notag \\
& \le 2  \lvert     \log ( f( F^{-1}(p )))  \lvert    \eu^{-2 (n+2)  \left(  \epsilon-\frac{p}{n+1} \right)^2 }+C_\epsilon^{(2)} O \left( \frac{1}{\sqrt{n}} \right) \notag\\
&\quad +2 C_q \left( \| f \|_{r+1}\right )^{\frac{r+1}{r}}  n^{\frac{1}{2}\left(1 -\frac{1}{q}\right)}  \eu^{- 2(n+2) \left( \epsilon- \frac{|(q-2) p-q +1|}{q(n-1)+2}  \right)^2  },
\end{align*} 
where the last inequality follows by bounding $\mathbb{P} \left[ \mathcal{A}^c \right] $ using  Lemma~\ref{lem:UBPAC}. This concludes the proof of Lemma~\ref{lem:Log_Term_bound}.


\section{Proof of the Bound in~\eqref{eq:ConstIntermStep1}} 
\label{app:eq:Stirling_bounds_on_c}
Recall from the proof of Lemma~\ref{lem:Log_Term_bound} that $\alpha := np$ and $\beta := n+1-np$ while $\alpha^*=q (\alpha-1)+1$, and  $\beta^*=q (\beta-1)+1$ for some $q > 1$. 
Thus, with reference to~\eqref{eq:ConstIntermStep1}, we want to show
that
\begin{align}
\label{eq:ConstIntermStep}
\frac{\Gamma(\alpha^*)^{ \frac{1}{q}}  \Gamma(\beta^*)^{ \frac{1}{q}} \Gamma(\alpha + \beta) }{\Gamma(\alpha^* + \beta^*)^{ \frac{1}{q}} \Gamma(\alpha)  \Gamma(\beta)} \leq C_q n^{\frac{1}{2}\left(1 -\frac{1}{q}\right)},
\end{align} 
where $C_q := {\rm{e}}^{1+\frac{2}{q}} \left(\sqrt{2 \pi }\right)^{\frac{1}{q}-1}  q^{- \frac{1}{2q}}$.

First, using that $ \Gamma(n+1)  = n!$, recall Stirling's approximation~\cite{abramowitz1970handbook}, which tells us that for $n \geq 1$,
\begin{align}
\label{eq:Stirling}
 \Gamma(n+1) = \kappa_n  \sqrt{2 \pi} n^{n+ \frac{1}{2}} \eu^{-n },
\end{align} 
where $ 1 \le  \eu^{\frac{1}{12n+1}} \le \kappa_n \le \eu^{\frac{1}{12n}} \le \eu$.

{
Using the bound in~\eqref{eq:Stirling}, we find
\[\Gamma( \alpha^*) = \Gamma(q(\alpha-1)+1) \leq  \eu  \sqrt{2 \pi} (q(\alpha-1))^{q(\alpha-1)+ \frac{1}{2}} \eu^{-q(\alpha-1) }, \]
and
\[\Gamma( \alpha) \geq  \sqrt{2 \pi} (\alpha-1)^{\alpha- \frac{1}{2}} \eu^{-(\alpha-1)}; \]
hence, by simplifying, we obtain
\begin{align}
\frac{\Gamma^{ \frac{1}{q}}( \alpha^*)}{\Gamma( \alpha)} 
& \leq \frac{\left({\rm{e}} \sqrt{2 \pi} (q(\alpha-1))^{q(\alpha-1)+\frac{1}{2}} {\rm{e}}^{-q(\alpha-1)}\right)^{\frac{1}{q}}}{\sqrt{2 \pi} (\alpha-1)^{\alpha-\frac{1}{2}} {\rm{e}}^{-(\alpha-1)}} \notag
\\& = {\rm{e}}^{\frac{1}{q}}\left(\sqrt{2 \pi }\right)^{\frac{1}{q}-1} (\alpha-1)^{-\frac{1}{2} + \frac{1}{2q}} q^{\alpha-1+\frac{1}{2q}}.
\label{eq:First_Stirling_approaximation_of_ratioMC_alpha}
\end{align}
We can similarly show that
\begin{align}
\frac{\Gamma^{ \frac{1}{q}}( \beta^*)}{\Gamma( \beta)} & \leq {\rm{e}}^{\frac{1}{q}}\left(\sqrt{2 \pi }\right)^{\frac{1}{q}-1} (\beta-1)^{-\frac{1}{2} + \frac{1}{2q}} q^{\beta-1+\frac{1}{2q}}.
\label{eq:First_Stirling_approaximation_of_ratioMC_beta}
\end{align}

Finally, we agin use the bound in \eqref{eq:Stirling} to find
\begin{align}
&\frac{\Gamma( \alpha+\beta)}{\Gamma^{ \frac{1}{q}}( \alpha^*+\beta^*)} = \frac{\Gamma( n+1)}{\Gamma^{ \frac{1}{q}}( q(n-1)+2)} \notag
\\& = \frac{n \Gamma(n)}{(q(n-1)+1)^{ \frac{1}{q}}  \Gamma^{ \frac{1}{q}}( q(n-1)+1)} \notag
\\& \leq \frac{\sqrt{2 \pi} {\rm{e}}^{-(n-2)} n (n-1)^{n-\frac{1}{2}}}{(q(n-1)+1)^{ \frac{1}{q}} \left(\sqrt{2 \pi} (q(n-1))^{q(n-1)+\frac{1}{2}} {\rm{e}}^{-q(n-1)} \right)^{\frac{1}{q}}} \notag
\\& = {\rm{e}}  \left(\sqrt{2 \pi }\right)^{1-\frac{1}{q}} n (n-1)^{\frac{1}{2}-\frac{1}{2q}} q^{-(n-1)-\frac{1}{2q}}(q(n-1)+1)^{-\frac{1}{q}} \nonumber
\\& \leq {\rm{e}}  \left(\sqrt{2 \pi }\right)^{1-\frac{1}{q}} n^{\frac{3}{2}\left(1-\frac{1}{q}\right)} q^{-(n-1)-\frac{3}{2q}}.
\label{eq:Second_Stirling_approaximation_of_ratioMC}
\end{align}
Combining~\eqref{eq:First_Stirling_approaximation_of_ratioMC_alpha}, \eqref{eq:First_Stirling_approaximation_of_ratioMC_beta}, and~\eqref{eq:Second_Stirling_approaximation_of_ratioMC} we arrive at
\begin{align*}
\frac{\Gamma^{ \frac{1}{q}}( \alpha^*)}{\Gamma( \alpha)} &\times \frac{\Gamma^{ \frac{1}{q}}( \beta^*)}{\Gamma( \beta)}  \times \frac{\Gamma( \alpha+\beta)}{\Gamma^{ \frac{1}{q}}( \alpha^*+\beta^*)}
\\& \leq {\rm{e}}^{1+\frac{2}{q}} \left(\sqrt{2 \pi }\right)^{\frac{1}{q}-1} n^{\frac{3}{2}\left(1-\frac{1}{q}\right)} (np-1)^{-\frac{1}{2}\left(1-\frac{1}{q}\right)}
\\& \qquad  (n(1-p))^{-\frac{1}{2}\left(1 - \frac{1}{q}\right)} q^{-\frac{1}{2q}}
\\& \leq {\rm{e}}^{1+\frac{2}{q}} \left(\sqrt{2 \pi }\right)^{\frac{1}{q}-1} n^{\frac{1}{2}\left(1 - \frac{1}{q}\right)} q^{- \frac{1}{2q}},
\end{align*}
which concludes the proof.
}

\section{Examples of Section~\ref{sec:DiscConcl}}
\label{app:examples}
%
%
%

\subsection{Density in~\eqref{eq:Den2}} 
For the pdf 
\begin{align*}
f_1(x)&= \frac{2}{x \log^3(x)},  \quad x\in ( \eu, \infty),
\end{align*}
the cdf and the quantile function are given by 
\begin{align*}
F(x) &=1-  \frac{1}{\log^2(x)},  \quad x\in ( \eu, \infty)\\
F^{-1}(p) &=\eu^{ \frac{1}{\sqrt{1-p}}},   \quad p \in (0,1). 
\end{align*}
{We note that $f_1$ satisfies conditions 1) and 2) in Theorem~\ref{thm:EntroVers}; hence, $K_1=O ({1}/{n})$ by Lemma~\ref{lem:uniform_entropy}
and $K_3=O ({1}/{\sqrt{n}})$ by Lemma~\ref{lem:Log_Term_bound}. However, $f_1$ does not satisfy condition 3) in Theorem~\ref{thm:EntroVers}. As we next show, for $f_1$ we indeed have $K_2=\infty$; hence, $D( X_{(np)}  \| G_{n,p}) = \infty$. This suggests that condition 3) (or a variation of it) in Theorem~\ref{thm:EntroVers} might indeed be necessary for convergence.}
First, note that 
\begin{align*}
 \E \left[  \frac{1}{ (1- U_{(k)})^\frac{m}{2} } \right]= c_{k,n} \int_0^1  \frac{u^{k-1} (1-u)^{n-k}}{ (1- u)^\frac{m}{2} } {\rm d} u,
 \end{align*}
 where $c_{k,n} = \frac{\Gamma(n+1)}{\Gamma(k) \Gamma(n+1-k)}$; hence,
\begin{align}
 \E \left[  \frac{1}{ (1- U_{(k)})^\frac{m}{2} } \right]=\infty,   \quad \frac{m}{2} \geq  n-k+1.\label{eq:Moment_is_infiinte}
 \end{align} 
Now, by using the Taylor expansion of $\eu^{x}$, we have that 
\begin{align}
 \E[ X_{(k)}]=  \E \left[ F^{-1}(U_{(k)}) \right] &=\E \left[ \eu^{(1-U_{(k)})^{-1/2}} \right] \notag \\
  &=\E \left[ \sum_{m=0}^\infty \frac{1}{m!} \frac{1}{ (1- U_{(k)})^\frac{m}{2} } \right] \notag \\
    &=\! \!\sum_{m=0}^\infty \!\frac{1}{m!}  \E \left[  \frac{1}{ (1- U_{(k)})^\frac{m}{2} } \right],\label{eq:afterTaylorSeries}
\end{align} 
where in the last step we have used  Tonelli's theorem (which holds even if the series diverges). 
Combining~\eqref{eq:Moment_is_infiinte} and~\eqref{eq:afterTaylorSeries}, we conclude that for every $k$ there exists an $m$ such that $\frac{m}{2} \geq  n-k+1$; hence, $ \E[ X_{(k)}]=\infty$.
This implies that 
\begin{align*}
\E \left[ \left( F^{-1} (U_{(np)}) -  F^{-1} (p)   \right)^2 \right] & = \E \left[ \left( X_{(np)}  -  \eu^{ \frac{1}{\sqrt{1-p}}}   \right)^2 \right]  \\
& = \infty,
\end{align*} 
which leads to $K_2=\infty$.

\subsection{Density in~\eqref{eq:Den1}}
For the pdf 
\begin{align*}
f_2(x)&= \frac{1}{x \log^2(x)},  \quad x\in (0, \eu^{-1}),
\end{align*}
the cdf and the quantile function are given by 
\begin{align*}
F(x)&=-  \frac{1}{\log(x)},    \quad x\in (0, \eu^{-1}),\\
F^{-1}(p)& =\eu^{-\frac{1}{p}},  \quad p \in (0,1).
\end{align*}
We note that $f_2$ satisfies conditions 2) and 3) in Theorem~\ref{thm:EntroVers}; hence, $K_1=O ({1}/{n})$ by Lemma~\ref{lem:uniform_entropy}
and $K_2=O ({1}/{\sqrt{n}})$ by Corollary~\ref{cor:T2asymptotic}. However, $f_2$ does not satisfy condition 1) in Theorem~\ref{thm:EntroVers}. Nevertheless, as we next show, we can still prove the convergence of $K_3$. We start by noting that
\begin{align*}
f(F^{-1}(p))= p^2 \eu^{ \frac{1}{p}};
\end{align*} 
hence, from~\eqref{eq:K3} we obtain
\begin{align*}
K_3&=     \E \left[ \log \left( \frac{f( F^{-1} (U_{(np)}) )}{f ( F^{-1} (p) )}\right)  \right]
\\& = 2 \E[ \log  U_{(np)} ]+ \E \left[ \frac{1}{U_{(np)}} \right ] -2\log(p)-\frac{1}{p}
\\& = 2 \left( \psi(np ) - \psi(n+1) \right)+\frac{n}{pn-1} -2\log(p)-\frac{1}{p},
\end{align*}
where $\psi$ is the digamma function.
By noting that the digamma function can be approximated as $\psi(x)= \log(x)-\frac{1}{2x} +O( \frac{1}{x^2})$  and also using the fact that $\psi(x+1)=\psi(x)+\frac{1}{x}$, we arrive~at
\begin{align*}
K_3 = \frac{n+1-pn-p^2n+p}{pn(pn-1)} + O \left( \frac{1}{n^2} \right),
\end{align*}
which,  together with  $K_1=O ({1}/{n})$
and $K_2=O ({1}/{\sqrt{n}})$, implies $D( X_{(np)}  \| G_{n,p}) =\Theta \left( \frac{1}{n} \right)$.

 \end{appendices} 
 
\bibliography{refs}
\bibliographystyle{IEEEtran}
\end{document}